\documentclass[aps,prl,showpacs,twocolumn]{revtex4}

\usepackage{psfig}
\usepackage{amssymb}

\begin{document}

\title{Fano-type interpretation of red shifts and red tails \\
in hole array transmission spectra}
\author{C. Genet, M.P. van Exter, and J.P. Woerdman}

\affiliation{Leiden University, Huygens Laboratory, P.O.Box 9504,
2300 RA Leiden, The Netherlands \\ {\sc Optics Communications {\bf
225}, 331 (2003)}}

\begin{abstract}
We present a unifying point of view which allows to understand
spectral features reported in recent experiments with
two-dimensional arrays of subwavelength holes in metal films. We
develop a Fano analysis of the related scattering problem by
distinguishing two interfering contributions to the transmission
process, namely a non-resonant contribution (direct scattering)
and a resonant contribution (surface plasmon excitation). The
introduction of a coupling strength between these two
contributions naturally induces resonance shifts and asymmetry of
profiles which satisfy simple scaling relations. We also report an
experiment to confirm this analysis.

\end{abstract}


\maketitle

\begin{figure}[tbh]
\centerline{\psfig{figure=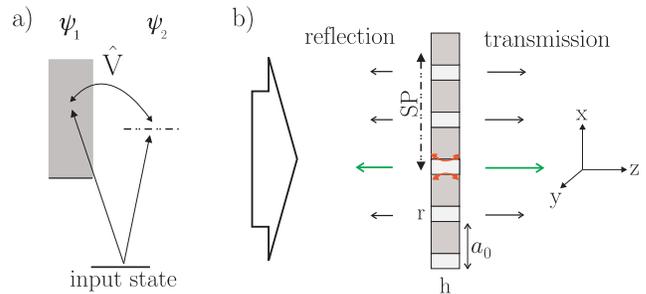,width=9cm}} \caption{a)
Formal representation of the Fano model for coupled channels. b)
Physical picture of the scattering process through the hole array
directly (straight arrows) or via SP excitation. The curved arrows
inside the hole represent the guided field. The period of the
array is denoted $a_{0}$, its thickness $h$ and the radius of each
hole $r$.} \label{fanofig}
\end{figure}

\begin{figure}[tbh]
\centerline{\psfig{figure=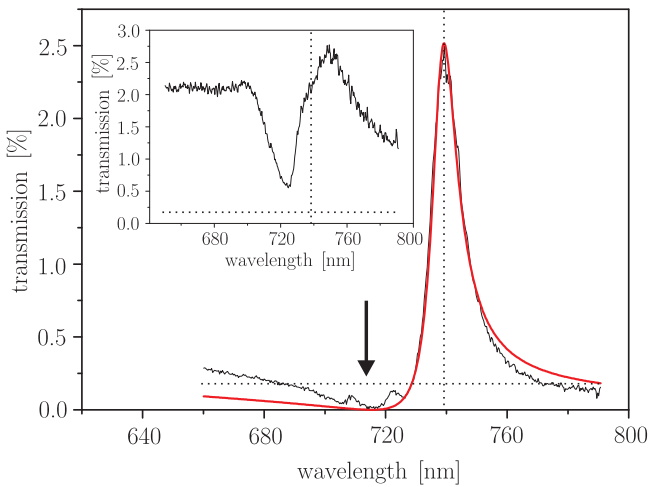,width=9 cm}}
\caption{Experimental transmission spectrum of the air-metal
($1,0$) SP resonance, obtained with an Au film of thickness $200$
nm with $a_{0}=700$ nm and $r=70$ nm. The smooth curve shows the
fitted Fano profile (\ref{fanoprofile}) by adjusting freely the
natural variables of (\ref{varnat}). The vertical dotted line
marks the position of the peak whereas the arrow shows the
position of the resonance given by equation
(\ref{plasmonrestheo}). The horizontal dotted line shows the Fano
background level. Inset: Experimental transmission spectrum for
the same array, but now in a film only $100$ nm thick. The former
dotted ``cross'' has been superimposed, emphasizing the background
increase and the dispersive line shape.} \label{spectrum}
\end{figure}

The observation of extraordinary transmission of a metal film
perforated with a two-dimensional periodic array of subwavelength
holes \cite{EbbesenNature1998,AltewischerNature2002} has been
followed by a considerable amount of theoretical work
\cite{GhaemiPRB1998,SalomonPRL2001,KrishnanOptCom2001,PopovPRB2000,MartinMorenoPRL2001}.
This work has not yet reached the maturity for a full
understanding of the precise mechanisms involved in this
transmission process. Important global characteristics of the
spectra have not yet been addressed satisfactorily: we refer here
to the fact that, in comparison with the naive ``band structure''
of surface plasmons (see equation (\ref{plasmonrestheo}) below),
resonances are red shifted and line shapes are asymmetric
\cite{GhaemiPRB1998,SalomonPRL2001,KrishnanOptCom2001}. In this
paper, we will address these issues by developing a unifying model
along the lines of the original Fano analysis of the
autoionization phenomenon in atomic physics \cite{Fano1961} and by
experimental confirmation of the essence of the model. We will
show that Fano's treatment yields useful insight in the problem of
transmission of light through metallic hole arrays by
distinguishing two interfering contributions to the transmission
process, namely a non-resonant contribution and a resonant one.
The latter is associated with surface plasmon (SP) excitations at
the interfaces of the hole array. In order to simplify the
physics, we allow for resonant SP excitation at one of the
interfaces of the hole array only. This assumption is
approximately valid for asymmetric arrays, i.e. arrays sandwiched
between two different dielectric media (such as air and glass) in
which a SP mode can be tuned into resonance on either the one or
the other interface, but not on both at the same time. Fabry-Perot
mediated coupling between two identical SP's living on separate
interfaces \cite{MartinMorenoPRL2001} can induce extra effects in
the symmetric case which we will not consider.\\
\indent The situation studied by Fano is one where the scattering
from an input state can take place either directly towards a
continuum of states (the scattered states) or via a quasibound
state, that is through some type of resonant state, which is then
coupled to the scattered states. This process therefore defines
two distinct types of scattering channels: one open channel
$\psi_{1}$ referring only to the continuum of states and one
closed channel $\psi_{2}$ with the resonant state, coupled to the
open channel, as symbolized in figure \ref{fanofig}a. A transition
from the input state straight to the open channel $\psi_{1}$ will
be called `direct' or `non-resonant' as opposed to the other
transition path that goes first through the quasibound state of
the closed channel $\psi_{2}$ before being scattered and is
therefore named `resonant'. The transition amplitudes associated
with each path will interfere
\noindent to define the total transition probability. This
situation leads to typical asymmetric resonance line shapes, named
(accordingly) as Fano profiles. We have to emphasize that Fano
analysis only addresses the global characteristics of the line
shape of the spectrum through ``structure parameters'' but does
not give theoretical expressions for these parameters, until a
detailed model is defined.\\
\indent The process of transmission through metallic hole arrays
is fundamentally a wave optics problem for which, in principle, a
full electromagnetic approach could be given. However, this
situation is, at the same time, a scattering problem which can be
mapped onto a formal quantum Hamiltonian problem, as soon as the
various scattering channels are identified. The scattering point
of view relates the transition amplitudes that are derived from
the Hamiltonian problem to the electromagnetic field scattering amplitudes.\\
\indent To calculate the total transition probability, one has to
solve the coupled channel problem which is defined by the
Hamiltonian
\begin{eqnarray}
\hat{H}=\hat{H}_{1}+\hat{H}_{2}+\hat{V}.  \label{hamiltonian}
\end{eqnarray}
The eigenstates of $\hat{H}_{1}$ correspond to the continuum
states of channel $\psi_{1}$ whereas $\hat{H}_{2}$ has one
discrete state corresponding to the resonant state in channel
$\psi_{2}$. The two channels are coupled via the coupling term
$\hat{V}$. Starting from an input state, the derivation of the
eigenstates of the full Hamiltonian $\hat{H}$ will give us the
transition amplitudes from the input state to the new eigenstates
of the coupled channel problem, taking into account the two
possible paths.\\
\indent The first path is non-resonant and simply related to the
direct scattering of the field through the subwavelength holes.
This contribution will be referred to as Bethe's contribution,
based on Bethe's solution to the problem of direct transmission of
the electromagnetic field through an infinitely thin ideal metal
with subwavelength holes \cite{Bethe1944}. We will simply adjust
this solution to our situation of holes drilled on a real metallic
film of finite thickness through a best-fit procedure, thus taking
into account in a effective way the influence of these realistic
conditions. This non-resonant path will correspond to the
$\hat{H}_{1}$ Hamiltonian formally describing a one-dimensional
scattering problem, taken along the $z$ direction, as shown in
figure \ref{fanofig}. Scattering states $\Psi_{E}^{p}$ will be
considered as eigenstates of $\hat{H}_{1}$, i.e.
$\hat{H}_{1}\Psi_{E}^{p} = E\Psi_{E}^{p}$. They will be defined
according to the direction $p$ of the incident plane wave and are
given asymptotically far away from the grating. The corresponding
scattering amplitudes, related to Bethe's solution, can be
gathered into a $S$-matrix $S_{B}$, which is symmetric
($S_{B}^{\dagger}=S_{B}$) due to transmission reciprocity property
of the array. These states are non-resonant states (defined in the
open channel) and belong as free fields to a two-folded continuum:
one transmission continuum and one reflection continuum. They are
detected as a background in the reflection and transmission
spectra. From the quantum point of view, these two continua,
specific of an electromagnetic scattering
process, are distinguished by the quantum number $p$.\\
\indent The second path corresponds to a resonant contribution,
related to the existence of discrete SP excitations at the
illuminated interface of the array. These excitations will be
considered as discrete eigenstates $\Psi_{sp}$ of the
$\hat{H}_{2}$ hamiltonian, i.e. $\hat{H}_{2}\Psi_{sp} =
E_{sp}\Psi_{sp}$. It is well known that these excitations
correspond to SP resonances defined on the periodic array by
wavevector matching, making use of the array momentum wavevector
\cite{RaetherBook}. At normal incidence, such resonances are, to
first order, given by complex resonance frequencies
\begin{eqnarray}
\hat{\omega}_{sp}=\left(n^{2}+m^{2}\right)^{\frac{1}{2}}\frac{2\pi
c}{a_{0}}
\sqrt{\frac{\varepsilon_{1}+\varepsilon_{2}}{\varepsilon_{1}\varepsilon_{2}}},
\label{plasmonrestheo}
\end{eqnarray}
where ($n,m$) are integers, $a_{0}$ is the period of the array and
$\varepsilon_{1}$ and $\varepsilon_{2}$ are respectively the
permittivities of the adjacent medium and the metal. The negative
imaginary part of $\hat{\omega}_{sp}$ corresponds to the internal
(non-radiative) damping of surface plasmon on a smooth interface.
This damping represents the coupling between the surface plasmon
and metal absorption losses. The real part of $\hat{\omega}_{sp}$
corresponds to the resonance frequencies. In practice,
transmission spectra of hole arrays peak at frequencies that are
typically $4\%$ smaller than the value given by the real part of
equation (\ref{plasmonrestheo}). This discrepancy, apparent in
many experimental works
\cite{GhaemiPRB1998,SalomonPRL2001,KrishnanOptCom2001}, is well
known though
basically not understood so far.\\
\indent Due to the array periodicity, the incident field can be
converted into a surface plasmon which will eventually be
scattered (reflected or transmitted) by reciprocally coupling
again with the array. This coupling between the discrete resonance
and the scattering states is formally given by the $\hat{V}$
operator (see also figure \ref{fanofig}b) which defines the
radiative damping of plasmons at the surface of the array,
considering the two scattering continua: a radiative transfer
directly to free space (to the reflection continuum) and a
radiative transfer through the holes (to the transmission
continuum). One can give interpretations for these radiative
dampings but the key result of this paper relies only on the
fundamental idea of the coupling with these two radiative
channels. Any interpretation of this coupling, aiming at
specifying the mechanisms of the radiative transfers, will not
question this result; this is the inherent strength of a Fano-type
analysis.  \\
\indent Fano has solved the coupled channel problem giving the
normalized eigenstate $\Phi_{E}$ of the full Hamiltonian $\hat{H}$
\cite{Fano1961}. The coupling between the resonant state
$\Psi_{sp}$ (SP excitation) and the scattering state
$\Psi_{E}^{p}$,
\begin{eqnarray}
\left\langle\Psi_{sp}\left|\hat{V}\right|\Psi_{E}^{p}\right\rangle
&=& V_{E}^{p},
\end{eqnarray}
is assumed hermitian. It introduces resonance shifts ($\Delta$)
and linewidths ($\Gamma$) whose expressions can be framed into a
Kramers-Kronig relation
\begin{eqnarray}
\Gamma_{E}=\frac{2\pi}{\hbar}\sum_{p}\left|V_{E}^{p}\right|^{2} \
\ , \ \ \Delta_{E}=\frac{1}{2\pi}{\mathcal P}\int{\rm
d}E^{\prime}\frac{\Gamma_{E^{\prime}}}{E-E^{\prime}}.
\label{kramerskronig}
\end{eqnarray}
The scattering dynamics of our system are fully contained in the
$S$-matrix element corresponding to the global transmission
amplitude
\begin{eqnarray}
t\equiv\left\langle\Phi_{E}\left|S\right|i\right\rangle,
\label{Smatrix}
\end{eqnarray}
with $\left|i\right\rangle$ denoting the input state, the
left-incident field. This expression takes into account direct
scattering from the input state to the continuum of scattering
states $\psi_{1}$ (Bethe contribution), specified through the
$S$-matrix $S_{B}$, as well as indirect scattering through the
resonant state of the closed channel $\psi_{2}$. The expression
for the $S$-matrix element (\ref{Smatrix}) takes the shape of a
sum of interfering terms from which the expression for the
transmission coefficient (transition probability) immediately
follows
\begin{eqnarray}
T=\left|t\right|^{2}
=\left|t_{B}\right|^{2}\frac{\left[E-\left(E_{sp}+\hbar\Delta\right)+\hbar\delta\right]^{2}}
{\left[E-\left(E_{sp}+\hbar\Delta\right)\right]^{2}+\left[\hbar\Gamma/2\right]^{2}},
\label{fanoprofile}
\end{eqnarray}
where $E=\hbar\omega$. The $\delta$ parameter defines the ratio
between the resonant transition amplitude and the direct
(background) transition amplitude \cite{Fano1961}. This expression
for the global transmission coefficient reveals the superposition
of the non-resonant `Bethe' contribution, with the transmission
coefficient $\left|t_{B}\right|^{2}$, and of the resonant part,
with a specific structure that turns out to be exactly a Fano-type
profile when written in the natural variables \cite{Fano1961}
\begin{eqnarray}
\epsilon=\frac{E-\left(E_{sp}+\hbar\Delta\right)}{\hbar\Gamma/2} \
\ , \ \  q=\frac{2\delta}{\Gamma}. \label{varnat}
\end{eqnarray}
The dimensionless $q$ parameter determines the asymmetry of the
profile and its sign the corresponding blue ($q >0$) or red ($q<
0$) tail \cite{note1}.\\
\indent Our analysis shows that the line shapes are fully
determined by 4 structure parameters: the resonance energy
$E_{sp}$, identified here as the real part of
$\hbar\hat{\omega}_{sp}$, the linewidth $\Gamma$, the resonance
shift $\Delta$ and the asymmetry $q$. The linewidth $\Gamma$
includes the plasmon damping terms, i.e. (i) radiative damping,
defined by the coupling strength between the surface plasmon
discrete state and the scattering states (reflection and
transmission continua), and (ii) non-radiative damping defined by
the imaginary part of $\hbar\hat{\omega}_{sp}$. The non-radiative
damping contribution is generally much weaker than the radiative
one, so that the resonance shift $\Delta$, given by the
Kramers-Kronig relation (\ref{kramerskronig}), is essentially due
to the additional radiation damping. This shift can explain the
discrepancy between expected resonances (see equation
(\ref{plasmonrestheo})) and actual peak positions. It defines
therefore non-perturbatively the modified SP's dispersion relation
when compared to the SP's dispersion relation on a smooth
interface, taking account of the ``dressed'' character of the SP
by coupling to free-field modes. Note that, even if it can easily
be parametrized in equations (\ref{fanoprofile},\ref{varnat}), our
one-resonance Fano description does not strictly allow for the
inclusion of SP-SP scattering since this would involve more than
one discrete state. However, this contribution is of higher order
in the limit taken here of small holes
as compared to the wavelength.\\
\indent A full electromagnetic theory would deduce this new
dispersion relation from the definition of the $S$-matrix of the
whole scattering process and from the extraction of its
corresponding poles \cite{PendryPRL1992,BarnesPRB1995}. The sign
of the asymmetry $q$ can be read off directly from the tail of an
experimental spectral profile of an isolated SP resonance.
Finally, the line shapes of transmission profiles show a
characteristic dip at the particular point $\epsilon =-q$
corresponding to a destructive interference effect between the two
channels. This destructive interference effect is concomitant to
the so called Wood anomaly
\cite{HesselOnlinerAO1965}.\\
\indent The definition of the structure parameters is directly
rooted in the introduction of the coupling between a non-resonant
and a resonant channel, as essentially shown by the Kramers-Kronig
relation (\ref{kramerskronig}). This relation provides us with
general relations between the damping and the resonance shift.
Although we have no explicit expressions for the coupling strength
$V_{E}^{p}$ between surface plasmons and scattering states, we can
argue generically how it depends on the ($\omega,r,h$)
parameters. \\
\indent With a simple interpretative model, it is easy to see that
the coupling strength is an increasing function of the frequency
$\omega$. Towards the reflection continuum, the coupling is
directly defined via the holes acting as Herz- or Rayleigh-type
dipole emitters induced by the longitudinally polarized surface
plasmon. Towards the transmission continuum, the (evanescent)
coupling through the holes is considered which is also an
increasing function of the frequency. This monotonic behaviour of
the coupling strength, that is of the linewidth $\Gamma$ (see
equation (\ref{kramerskronig})), implies, when inserted into this
Kramers-Kronig relation (\ref{kramerskronig}), that the resonance
shift $\Delta$ is negative, i.e. a red shift, as noticed by
several authors
\cite{GhaemiPRB1998,SalomonPRL2001,KrishnanOptCom2001}. In fact,
the absence of a detailed description for the linewidth-resonance
shift ratio forces us to consider $\Gamma$ as a free parameter,
including both the effects of radiative damping (surface plasmons
coupled to scattering states) and non-radiative damping. \\
\indent This coupling is also strengthened when, at fixed
frequency, the hole radius is increased. This lowers the cut-off
frequency of the hole waveguide and actually increases the
coupling of surface plasmons towards scattering states of the
reflection and transmission continua, as the radiating dipoles get
stronger. The linewidth is therefore predicted to be an increasing
function of the hole radius ($\partial_{r}\Gamma> 0$) and, at the
same time, the red shift of the resonance should be more
pronounced ($\partial_{r}\left|\Delta\right|> 0$). These
variations have been observed experimentally
\cite{SalomonPRL2001}.\\
\indent If the thickness $h$ of the array is increased, the
coupling towards the transmission continuum decreases. In the
hypothetic limit of infinite thickness, only radiation damping on
the reflection interface remains. As $h$ is increased, the
progressive disappearance of the radiative channel through the
holes, due to higher (non-radiative) losses inside the waveguide,
induces a narrowing of the linewidth ($\partial_{h}\Gamma< 0$) and
consequently a reduced red shift
($\partial_{h}\left|\Delta\right|< 0$) of resonance. Two regimes
of variation for linewidths and resonance shifts can be understood
from the introduction of a typical thickness above which the
radiative coupling is mainly towards the reflection continuum. The
variations $\partial_{h}\Gamma< 0$ and
$\partial_{h}\left|\Delta\right|< 0$, along with the existence of
a critical thickness, are again in agreement with recent
measurements \cite{DegironAPL2002}.\\
\indent Figure \ref{spectrum} shows measurement transmission data
from one of our experiments, for the $(1,0)$ SP resonance on the
air-metal side of our array. This array is made of a $200$ nm
thick gold (Au) film perforated with a square lattice of $70$ nm
radius holes spaced with a $700$ nm lattice period. Our sample is
asymmetric: the Au film is evaporated on a glass substrate with a
$10$ nm thick bonding layer of titanium (Ti). Strong absorption in
this bonding layer prevents any surface plasmon from being excited
on the metal-glass interface and, as such, guarantees that only
one interface contributes, in agreement with the general frame of
our model. This profile has a wavelength red tail, i.e. $q$ is
negative \cite{note1}. We notice that the position of the measured
peak (at $\lambda=739$ nm) is $\sim 4\%$ red shifted when compared
to the position of the corresponding SP resonance
$\lambda_{sp}=712$ nm naively expected from equation
(\ref{plasmonrestheo}) with a period of the array $a_{0}=700$ nm.

We attribute this shift to a negative $\Delta$ parameter, as to be
expected from the above general arguments. It should be noticed
that the various parameters of equation (\ref{fanoprofile}) are
intertwined in their effects on the line shape. Shifts of SP
resonances are determined for one part by $\Delta$ and for the
other part by the asymmetry itself. In natural units, the peak of
the Fano profile is centered on $\epsilon=1/q$. A red tail
asymmetry ($q < 0$) induces therefore an extra red shift, though
small (less than $1\%$) when compared to the red shift that is
expected from $\Delta$. On the other hand, the non-resonant
transmission coefficient scales as
$\left|t_{B}\right|^{2}\sim\left(r/\lambda\right)^{4}$, according
to Bethe's theory \cite{Bethe1944}. This defines a non-resonant
background colored towards the blue. The Fano peak will therefore
be blue-shifted when superposed on this background. This blue
shift can again be neglected when compared to the red
shift $\Delta$.\\
\indent In figure \ref{spectrum}, we have superposed a fitting of
the experimental curve with a profile defined by equation
(\ref{fanoprofile}) \cite{note0}. The fitted parameters indicate a
red shift of $4\%$ determined by $\Delta <0$ and a relative
linewidth of about $3\%$, that is of the order of the shift. The
background is found to be only about $0.2\%$, which is consistent
with Bethe's calculations for this system. Below $710$ nm, the
tail of the
air-metal ($1,1$) resonance is emerging.\\
\indent As mentioned above, the value of the Fano asymmetry $q$
characterizes the ratio between the resonant transition amplitude
and the background transition amplitude. Our experimental curve
confirms this scaling with an asymmetry of the order of $4$ which
can also be deduced from the fitting parameters. As a check of the
Fano description, we have reduced the Au film thickness from $200$
nm to $100$ nm with the same $10$ nm thick Ti layer. With this
reduction, the film thickness is only a few times the Au skin
depth allowing another direct leakage channel. This should
increase the non-resonant background contribution and thus lower
the value of $q$. We have confirmed this experimentally: see the
inset of figure \ref{spectrum} where the background level is
increased and the profile of the resonance gets close to a
dispersive shape which corresponds to a lower $q$ value.\\
\indent In conclusion, we have shown that a Fano analysis is a
unifying tool to understand hole array spectra since simple
scaling laws can be inferred. Only one resonance has been
addressed in this paper and one should now extend this work to the
collection of SP excitations on the array, i.e. to multichannel
theory.

{\it Note added in proof - }During the process of publication of
this paper, we learned about the theoretical work of Sarrazin {\it
el al.} \cite{Sarrazin2003} which shows important overlapping with
our. Nevertheless, we put here the emphasis on the concept of
coupled scattering channels in the context of experimental
spectra, in order to understand corresponding lineshapes.

{\it Acknowledgments - }This work forms part of the program of
FOM; we would like to thank G. Nienhuis for useful discussions.


\begin{thebibliography}{}
\bibitem{note1} The transmission coefficient is measured as a
function of the wavelength. A negative asymmetry $q_{E} <0$ for
frequency spectra immediately corresponds to a positive asymmetry
$q_{\lambda}
>0$ for wavelength spectra.
\bibitem{note0} The two wiggles seen around the dip of the profile
are an experimental artefact due to non perfect alignment of the
incident beam and the normal axis of the array.
\end{thebibliography}
\end{document}